\documentclass{article}

\usepackage{epsf}
\usepackage{psfig}
\usepackage{amssymb,amsmath}

\begin{document}

\vspace*{-.6in} \thispagestyle{empty}
\begin{flushright}
CFP--2003--15\\
\end{flushright}
\vspace{.2in} {\Large
\begin{center}
{\bf Late--time Cosmic Dynamics from  M--theory}
\end{center}}
\vspace{.2in}
\begin{center}
Pedro G. Vieira$^{\dagger}$
\\
\vspace{.2in}
$^{\dagger}$\emph{Centro de F\'\i sica do Porto and Departamento de F\'\i sica,\\
Faculdade de Ci\^encias, Universidade do Porto\\
Rua Campo Alegre 687, 4169--007 Porto, Portugal}
\end{center}

\vspace{.3in}

\begin{abstract}
We consider the behaviour of the cosmological acceleration for time-dependent hyperbolic 
and flux compactifications of M-theory, with an exponential potential. 
For flat and closed cosmologies it is seen that a positive acceleration is always transient 
for {\em both} compactifications. For open cosmologies, {\em both} compactifications can give
at late times periods of positive acceleration. As a function of proper time this acceleration 
has a power law decay and can be either positive, negative or oscillatory. 
\end{abstract}

\newpage

According to current observations it is believed that the universe is now experiencing a period 
of recent positive acceleration. A natural question to ask is how generic is this 
acceleration, and therefore the associated dark energy, 
in the context of string and M--theory compactifications. In particular, there 
has been a considerable amount of recent work showing that, provided the volume of the
compact space is time--dependent, periods of acceleration can indeed occur [3]--[17]. These findings
evaded the previous ``no--go theorem'' \cite{Gibbons,MaldNunez}, which did not allow for
such time--dependence.

Let us then consider a four--dimensional FRW spacetime, together with an internal space 
of dimension $n$. We shall assume that only the volume of the internal space depends
on the time coordinate. Then, both in flux and hyperbolic compactifications, the 
four-dimensional action reduces to
\begin{equation*}
S=\frac{1}{16\pi G}\int d^4x \sqrt{g} \left( R -\frac{1}{2}\left(\partial\psi\right)^2
-V(\psi)\right)\ ,
\end{equation*}
where $V(\psi)=\Lambda\,e^{-a\psi}$ and $\Lambda$ is a positive constant. 
This effective theory is thus defined by a family
of potentials parametrized by a constant $a$. For hyperbolic compactifications, with an internal
space of dimension $n\ge 2$, this constant is given by $a=\sqrt{(n+2)/n}$ and therefore
lies in the range
\begin{equation*}
1<a\le\sqrt{2}<\sqrt{3}\,.
\end{equation*}
On the other hand, for flux compactifications one has
\begin{equation*}
a\ge \sqrt{3}\,.
\end{equation*}
Regardless of initial conditions, the late--time evolution of these cosmologies, including 
the particular solutions studied recently, can be analysed with generality following the work 
of Halliwell \cite{Halliwell}. 

In this note, we shall identify, for the above supergravity compactifications, every possible asymptotic 
behaviour for the cosmic acceleration. The exact solution for flat universes was found, for any constant
$a$, by Townsend \cite{Townsend}. For both compactifications, we shall see that an accelerating
epoch is necessarily transient\footnote{The case $a=1$, which would lead to an eternally accelerating 
universe without a future event horizon \cite{Townsend}, is not included in the above compactifications.}. 
Similarly, for closed cosmologies, an accelerating phase must be transient. In the case of open
cosmologies, we shall see that one always has late--time periods of acceleration and/or
deceleration which decay with proper time as a power law. 

Let us start by reviewing Halliwell's work, which translates the above problem to that of finding 
the solutions of a two--dimensional dynamical system. This is done by introducing the 
lapse function $N\left(t\right)$ in the FRW metric:
\begin{equation*}
-N^{2}\left(t\right)\,dt^{2}
+e^{2A\left(t\right)}\,ds^{2}\left(\mathcal{M}_{k}\right)\,,
\end{equation*}
where $\mathcal{M}_{k}$ is $\mathbb{M}^{3}$, $S^{3}$ or $H^{3}$ 
depending on whether $k=0,1,-1$. Setting $N^{2}=V^{-1}$ and denoting
the derivatives with respect to $t$ by dots, the Friedmann equation becomes
\begin{equation*}
\dot{A}^{2}-\frac{1}{12}\,\dot{\psi}^{2}
=\frac{1}{6}-\frac{k}{\Lambda }\,e^{a\psi -2A}\,.
\end{equation*}
Therefore the hyperbola $\dot{A}^{2}=\frac{1}{12}\,\dot{\psi}^{2}+\frac{1}{6}$
divides the regions where $k$ is positive or negative. Moreover, the
equations of motion reduce to the following dynamical system
\begin{eqnarray*}
&&
\ddot{A}=\frac{1}{6}-\frac{1}{6}\,\dot{\psi}^{2}-\dot{A}^{2}
+\frac{a}{2}\,\dot{A}\,\dot{\psi}\,, 
\\
&&
\ddot{\psi}=\frac{a}{2}\,\dot{\psi}^{2}-3\dot{A}\,\dot{\psi}+a\,,
\end{eqnarray*}
which has fixed points in the $\dot{\psi}\dot{A}$--plane 
(neglecting the ones obtained by $t\rightarrow-t$, which correspond to a contracting universe) 
\begin{eqnarray}
P_{1} &=&\left(\frac{a\sqrt{2}}{\sqrt{3-a^{2}}}\,,\frac{1}{\sqrt{2(3-a^{2})}}\right) \,,  
\label{fixpoint} \\
P_{2} &=&\left(1,\,\frac{a}{2}\,\right) \,.  
\notag
\end{eqnarray}
Note that $P_{1}$ is always on the $k=0$ hyperbola. The attractor solution is $P_{1}$ for $0<a<1$
and $P_{2}$ otherwise. For $a>\sqrt{3}$ the point $P_{1}$ no longer exists.

Now that we wrote the basic equations derived by Halliwell, let us
analyse the behaviour of the acceleration. To check if the corresponding cosmological 
solution is accelerating, we must check the positivity of the second derivative of the scale 
factor $S$ with respect to proper time
\begin{equation}
\frac{d^2S}{d\tau^2} = 
\left( \frac{1}{N}\frac{d}{dt}\right)^{2}e^{A} = 
\frac{V\,e^{A}}{6}\left(1-{\dot{\psi}}^{2}\right)\,,
\label{accel}
\end{equation}
where in the last step we used the equations of motion. A positive acceleration
is therefore equivalent to
\begin{equation*}
\dot{\psi}^{2}<1\,.
\end{equation*}
It is convenient to define the quantity $w(t)$ by the equation of state for the
scalar pressure $p=w\rho$. Then, it is straightforward to show that the equations
of motion lead to
\begin{equation*}
w=\frac{\dot{\psi}^2-2}{\dot{\psi}^2+2}\ .
\end{equation*}
Hence vertical lines on the $\dot{\psi}\dot{A}$--plane are lines
of constant $w$. The $\dot{A}$--axis corresponds to  $w=-1$, and the stripe where the 
universe accelerates corresponds to $-1<w<-1/3$. As a function of $\dot{\psi}^2$, $w$ is
a growing function whose maximum is 1. 

\begin{figure}
\begin{picture}(0,0)(0,0)
\end{picture}
\centering\psfig{figure=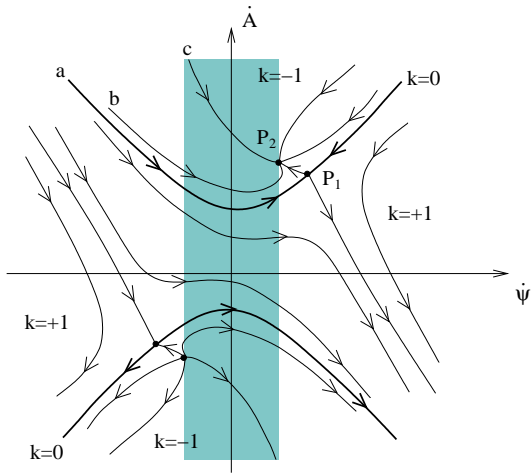,width=7cm} 
\caption{\small{Trajectories in the $\dot{\psi}\dot{A}$--plane for $1<a<\sqrt{4/3}$. When
$\sqrt{4/3}<a<\sqrt{3}$ the diagram is similar but the trajectories spiral around the 
stable attractors. These are the two regimes associated with hyperbolic compactifications.
Inside the shaded region the universe is accelerating, whereas outside it
decelerates.}}
\label{fig1}
\end{figure}

Let us analyse the various trajectories in the $\dot{\psi}\dot{A}$--plane drawn in 
\cite{Halliwell}, and check whether they correspond to an accelerating universe. 
Consider first the case of a closed universe. Then, for $a<1$ there are 
trajectories in the phase plane which exhibit late--time acceleration. These 
trajectories flow towards the attractor $P_1$ which is located inside
the acceleration stripe. As shown in \cite{Halliwell}, the scale factor
has the power law behaviour $S\propto \tau^{1/a^2}$ and therefore
the geometry has a future event horizon.
On the other hand, for $a\ge 1$, which includes M--theory compactifications, 
there is a runaway behaviour with a decelerating universe, as can be seen 
in both figures 1 and 2. 

\begin{figure}
\begin{picture}(0,0)(0,0)
\end{picture}
\centering\psfig{figure=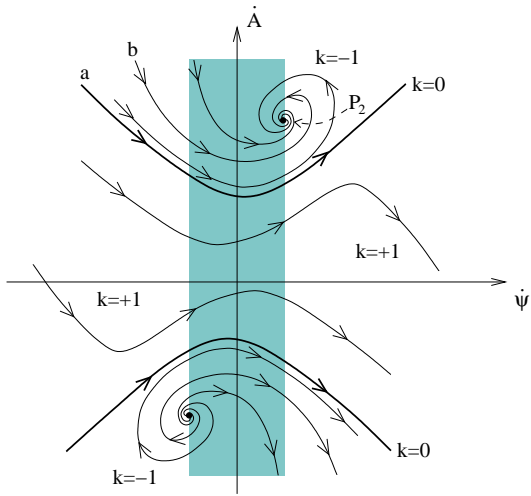,width=7cm} 
\caption{\small{Trajectories in the $\dot{\psi}\dot{A}$--plane for $a>\sqrt{3}$, corresponding
to flux compactifications. For open cosmologies, the late time behaviour of the acceleration is always 
oscillatory.}}
\label{fig2}
\end{figure}

Next, let us consider the fixed point $P_{1}$, by concentrating, for the
moment, on a flat universe with $k=0$. The class of solutions for a flat
universe and arbitrary constant $a$ were found explicitly in \cite{Townsend}.
For $0<a<\sqrt{3}$, the point $P_{1}$ is always an attractor, 
\textit{if we restrict to the }$k=0$ \textit{hyperbola}. For $a< 1$, and therefore not for
hyperbolic nor flux compactifications, the fixed point has $\dot{\psi}^{2}< 1$. 
Hence, the asymptotic solution is \textit{accelerating}. The case $a=1$ has a 
solution which accelerates and a solution which decelerates, depending if one starts
from $\dot{\psi}=\mp\infty$. In particular, the accelerating solution does not have 
a future event horizon \cite{Townsend}. Within the range $1<a<\sqrt{3}$,  which includes
hyperbolic compactifications, the asymptotic solution is 
always decelerating (figure \ref{fig1}). For example, 
following \cite{TownWohl}, consider the trajectory (a) in figure \ref{fig1}, starting
with large negative $\dot{\psi}$ (the field rolling up the potential). Then,
as $\dot{\psi}^{2}$ becomes less then $1$, we enter into a period of 
\textit{transient acceleration}, followed again by a period of deceleration.
Finally, for $a>\sqrt{3}$, the $k=0$ trajectory (a) of figure \ref{fig2},
associated to flux compactifications, has a runaway behaviour, going through a 
period of transient acceleration.

Let us move to the case of $k=-1$, with $a>1$, where there is always an
attractor at $P_{2}$. Note that $P_{2}$ is on the boundary of the region 
$\dot{\psi}^{2}<1$ of acceleration, because at late times curvature dominates and
one has linear expansion. Consequently, these geometries do not have a future
event horizon. The behaviour of the
trajectories converging to $P_{1}$ changes at $a=\sqrt{4/3}$ (see \cite{Halliwell}
for details). For $a<\sqrt{4/3}$ the trajectories are as in figure \ref{fig1},
whereas for $a>\sqrt{4/3}$ the trajectories spiral to $P_{2}$, like in
figure \ref{fig2}. Whenever $a<\sqrt{4/3}$, we can arrange initial conditions to
have a period of transient acceleration similar to the $k=0$ case, as in the
trajectory (b) in figure \ref{fig1}. Other initial conditions will
give a cosmology with a late time decaying acceleration, as for the
trajectory (c) in the same figure. When $a>\sqrt{4/3}$ (so, in particular, for all
flux compactifications and for hyperbolic compactifications with $n<6$), 
\textit{independently of initial conditions}, we have a cyclic
behaviour, with the acceleration oscillating around zero, with decreasing
magnitude. 

It is now clear that, for  flat universes, a positive acceleration is \textit{always}
transient for \textit{both} type of compactifications and one \textit{always} has
late--time deceleration. This is in contrast with
open cosmologies, where \textit{both} compactifications can give, at late times,
periods of positive acceleration. In the following we shall study the behaviour of this
late--time acceleration as a function of proper time.

Let us then analyse the dynamical system near the attractor $P_2$.
After the standard linearization, one obtains the asymptotic behaviour
\begin{equation}
\left(
\begin{array}{c}
\dot{\psi}\\
\dot{A}
\end{array}
\right)
=
\left(
\begin{array}{c}
1\\
a/2
\end{array}
\right)
+
\left(
\begin{array}{c}
V_1\\
V_2
\end{array}
\right)\,e^{\lambda_+t}
+
\left(
\begin{array}{c}
W_1\\
W_2
\end{array}
\right)\,e^{\lambda_-t}\ ,
\label{linear}
\end{equation}
where ${\bf V}$ and ${\bf W}$ are the eigenvectors, and the eigenvalues are
\begin{equation*}
\lambda_{\pm}=\frac{1}{2}\,\left(-a\pm\sqrt{\Delta}\right)\ ,
\ \ \ \ \ \ \ \ \ \ \ \ \ 
\Delta=4-3a^2\ .
\end{equation*}
It is then straightforward to determine the proper time as a function of the 
time coordinate $t$ which, to leading order, is 
\begin{equation}
\tau=\tau_0\,\exp\left(\,{\frac{a}{2}\,t}\,\right) + \cdots\ .
\label{time}
\end{equation}
Integrating equation (\ref{linear}) and replacing in the general formula for the 
acceleration (\ref{accel}), one obtains the following asymptotic behaviour as a function
of proper time
\begin{equation}
\frac{d^2S}{d\tau^2} = \frac{1}{\tau^2}\,\left(\alpha_+\,\tau^{\,\frac{\sqrt{\Delta}}{a}}
+\alpha_-\,\tau^{\,-\frac{\sqrt{\Delta}}{a}}\right)\ ,
\label{accel3}
\end{equation}
where $\alpha_\pm$ are integration constants.

Consider first $a\le\sqrt{4/3}$, corresponding to real eigenvalues. Then the
asymptotic behaviour for the acceleration reads
\begin{equation*}
\frac{d^2S}{d\tau^2}\propto\tau^{-2+\frac{\sqrt{\Delta}}{a}}\ .
\end{equation*}
On the other hand, for $a>\sqrt{4/3}$ the eigenvalues are complex, thus giving
rise to the oscillatory behaviour
\begin{equation*}
\frac{d^2S}{d\tau^2}\propto
\tau^{-2}\,\cos\left(\frac{\sqrt{-\Delta}}{a}\,\ln\tau+\varphi\right)\ ,
\end{equation*}
where $\varphi$ is a phase. Hence the acceleration will oscillate with an amplitude 
decreasing as $\tau^{-2}$, falling faster then in the case with $a\le\sqrt{4/3}$.
The periods of acceleration and deceleration become longer as $\tau\rightarrow\infty$.
In fact, if the acceleration vanishes at $\tau=\tau_1$, then the next zero will occur after an
interval
\begin{equation*}
\delta\tau = \tau_1 \left( e^\beta - 1\right)\ ,
\ \ \ \ \ \ \ \ \ \ \ \ \ \ 
\beta=\frac{\pi a}{\sqrt{-\Delta}}\ .
\end{equation*}
Thus periods of acceleration or deceleration grow linearly with cosmological time.

Finally, it is tempting to define an effective matter
and an effective cosmological constant associated with the scalar field by
writing the deceleration and total density parameters as
\begin{eqnarray*}
q=-\frac{S''S}{S'{}^2}=\frac{\Omega_M}{2}-\Omega_\Lambda\,,
\\
\Omega=1+\frac{k}{S'{}^2}=\Omega_M+\Omega_\Lambda\,,
\end{eqnarray*}
where primes denote derivatives with respect to proper time $\tau$.
Inverting this system of equations we conclude that lines of constant $\Omega_M$ are
straight lines and lines of constant $\Omega_\Lambda$ are ellipses, respectively
given by 
\begin{equation*}
6\,\Omega_M\dot{A}^2 = \dot{\psi}^2\ ,\ \ \ \ \ \ \ \ \
6\,\Omega_\Lambda\dot{A}^2 + \frac{\dot{\psi}^2}{2} = 1\ .
\end{equation*}
At the attractor, $P_2$, one has  $\Omega_M=2\Omega_\Lambda=2/3a^2$ while
at the saddle point, $P_1$, one has $\Omega_M=1-\Omega_\Lambda=2a^2/3$. The present acceleration
of the universe derived from supernovae measurements of the $q$ parameter, can then be generically 
reproduced by trajectories on their way to the attractor. For example, trajectories like (b) 
in both figures go through an accelerating phase compatible with supernovae observations. 
Also, it is clear
that the system evolves towards an attractor where $\Omega_M\sim\Omega_\Lambda$, therefore
it would not be surprising that both components were of the same order today. This fact could
explain the cosmic coincidence problem if the late time dynamics of the universe is determined by
the compactification scalar. Notice that, usually one thinks of the quintessence field as the source 
of dark energy only. Here one is naively assuming that both dark matter and dark energy are generated by the
quintessence field alone. While this interpretation seems to resolve the coincidence problem
it raises serious problems for small scale structure formation. In fact, even if it is possible to
have acceptable growth of baryonic perturbations, it turns out that the perturbations on the Newtonian potential 
caused by the coupling of the scalar field to gravity do not grow, and therefore
cannot explain the mysterious dark matter. Alternatively, we could study the evolution of this
model coupled to additional matter. In particular, it would be interesting to investigate
if it evolves to a period of eternal acceleration with
$\Omega_M\sim\Omega_\Lambda$.

\section*{Acknowledgments}
I would like to thank Lorenzo Cornalba and Miguel Costa for suggesting this problem
and for guidance. I would also like to thank  Carlos Herdeiro and Pedro Avelino
for helpful comments. 
This work is partially funded by CERN under contract POCTI/FNU/49507/2002--FEDER.

\end{document}